\documentclass[twocolumn,showpacs,preprintnumbers,aps,prb]{revtex4-1}
\usepackage{graphicx,color}
\usepackage{dcolumn}
\usepackage{bm}

\def\lsco{La$_{2-x}$Sr$_x$CuO$_4$}
\def\lbco{La$_{2-x}$Ba$_x$CuO$_4$}
\def\lsno{La$_{2-x}$Sr$_x$NiO$_4$}

\def\ybco{YBa$_2$Cu$_3$O$_{6+x}$}
\def\lbcoate{La$_{1.875}$Ba$_{0.125}$CuO$_{4}$}
\def\qaf{${\bf Q}_{\rm AF}$}
\def\ecross{$E_{\rm cross}$}

\begin{document}
\preprint{}

\title{Energy-dependent crossover from anisotropic to isotropic magnetic dispersion in lightly-doped La$_{1.96}$Sr$_{0.04}$CuO$_{4}$}
\author{M.~Matsuda}
\affiliation{Quantum Condensed Matter Division, Oak Ridge National Laboratory, Oak Ridge, Tennessee 37831, USA}

\author{M.~Fujita}
\affiliation{Institute for Materials Research, Tohoku University, Katahira, Sendai 980-8577, 
Japan}

\author{G. E. Granroth}
\affiliation{Quantum Condensed Matter Division, Oak Ridge National Laboratory, Oak Ridge, Tennessee 37831, USA}

\author{K.~Yamada}
\affiliation{Institute of Materials Structure Science, High Energy Accelerator Research Organization (KEK), Ibaraki 305-0801, Japan}

\author{J.~M.~Tranquada}
\affiliation{Condensed Matter Physics \&\ Materials Science Dept., Brookhaven National Laboratory, Upton, New York 11973, USA}
\date{\today}

\begin{abstract}
Inelastic neutron scattering experiments have been performed on lightly-doped La$_{1.96}$Sr$_{0.04}$CuO$_{4}$, which shows diagonal incommensurate spin correlations at low temperatures. We previously reported that this crystal, with a single orthorhombic domain, exhibits the ``hourglass" dispersion at low energies [Phys. Rev. Lett. 101, 197001 (2008)]. In this paper, we investigate in detail the energy evolution of the magnetic excitations up to 65 meV. It is found that the anisotropic excitations at low energies, dispersing only along the spin modulation direction, crossover to an isotropic, conical dispersion that resembles spin waves in the parent compound La$_2$CuO$_{4}$. The change from two-fold to full symmetry on crossing the waist of the hourglass reproduces behavior first identified in studies of underdoped \ybco.  We discuss the significance of these results.
\end{abstract}

\maketitle

\section{Introduction}

The nature of spin correlations and their relationship to superconductivity in layered cuprates remains controversial.  This is not due to a lack of experimental studies, as there has been considerable progress in characterizing the magnetic excitations\cite{birg06,fuji12a}; rather, the challenge is one of interpreting the results in the absence of a satisfactory model for itinerant antiferromagnetism in a strongly-correlated system.  A complementary problem is that of understanding the pseudogap phenomena observed in electronic spectroscopies.\cite{norm05}   In the absence of a fully predictive model, we can continue to explore and extend the experimentally-identified trends, and make comparisons with predictions of simplified models.

One of the established trends is the development of an ``hourglass'' dispersion in under- to optimally-doped cuprates.\cite{fuji12a}  As suggested by the name, low-energy and high-energy excitations disperse outwards from the energy \ecross\ characterizing the waist of the hourglass, with the dispersions centered about the antiferromagnetic wave vector \qaf.  A particularly interesting result has been observed in the \ybco\ (YBCO) family.  There the CuO$_2$ planes have an orthorhombic symmetry such that orthogonal Cu-O bonds are not equivalent.  Neutron scattering measurements on arrays of detwinned single crystals have revealed that the excitations below \ecross\ have only two-fold rotational symmetry about \qaf, whereas the excitations at energies above \ecross\ have at least four-fold rotational symmetry.\cite{stoc04,stoc05,hink07,hink10}  The reduced symmetry below \ecross\ has been associated with the concept of nematic electronic correlations.\cite{kive98}  At the same time, the downward dispersion in \ybco\ with $x\gtrsim0.5$ is only resolved at temperatures below the superconducting transition $T_c$,\cite{bour00,hink10} which has encouraged interpretations that it is associated with the spin-resonance feature of the superconducting state.\cite{esch06}  Note, however, that nematic anisotropy of low-energy spin excitations in crystals with $x\le 0.45$ has been mapped to $T\gg T_c$.\cite{haug10}

In the \lsco\ (LSCO) and \lbco\ (LBCO) systems, the dispersion of excitations below \ecross\ is readily observed at $T>T_c$,\cite{aepp97,chri04,lips09,fuji04,xu07} and in the case of \lbcoate, in particular, the excitations connect to incommensurate magnetic superlattice peaks.\cite{fuji04,tran04}  The latter point provides motivation to associate the dispersion with excitations about a stripe-ordered state.\cite{tran95a,kive03,vojt09,zaan01}  Unfortunately, it has not been practical to test the rotational symmetry of the excitations in superconducting LSCO or LBCO due to constraints of crystal symmetry.  Such a measurement would be valuable in testing theoretical models.  Starting from a state with charge and spin stripe order, the simplest sort of model to consider involves the magnetic moments only and ignores the charge carriers, treating the system in terms of spin ladders with a weakened coupling across the charge stripes.  The anisotropy of the striped ground state imprints itself in the magnetic dispersions, showing up at all energies.\cite{tran04, krug03,carl04,koni08,grei10}
Averaging over twinned stripe domains\cite{tran04,koni08,grei10} or allowing for disorder\cite{vojt06} can restore the four-fold symmetry at high energy, but also forces its presence at low energy.  A calculation based on the time-dependent Gutzwiller approximation applied to the Hubbard model  does slightly better, but still exhibits two-fold symmetry at all energies.\cite{seib06,seib12a}    Hence, stripe-based calculations thus far have not been able to describe the combination of two-fold and four-fold symmetry observed in \ybco.\cite{stoc04,stoc05,hink10,haug10}  This leaves us with the question of whether the symmetry recovery in YBCO is incompatible with stripe correlations or simply with the stripe models considered so far.

An opportunity for testing the anisotropy of magnetic dispersions in \lsco\ occurs in the ``spin-glass'' regime, $0.02\lesssim x \lesssim0.055$.    Elastic neutron scattering measurements have established that the incommensurate spin modulation, oriented along the diagonal of a CuO$_2$ plaquette for this doping, has a unique orientation with respect to the orthorhombic axes of the crystal lattice.\cite{waki00,fuji02c}  Note that the orthorhombic axes in this case are along the plaquette diagonals, with $a < b \approx 5.4$~\AA, leaving the orthogonal Cu-O bond directions equivalent.  We will specify wave vectors in terms of the orthorhombic cell, so that ${\bf Q}_{\rm AF} = (1,0,0)$.  The incommensurate modulation is along the [010] direction.

By performing inelastic neutron scattering measurements on a crystal with a single orthorhombic domain, it is possible to test the character of the spin excitations.  In previous experiments using a triple-axis spectrometer to measure nearly-single-domain crystals, we have shown for LSCO $x=0.04$ and 0.025 that the low-energy spin fluctuations ($\lesssim10$~meV) exhibit an anisotropic dispersion inwards towards \qaf\ from the elastic incommensurate peaks.\cite{mats08,mats11}  Measurements at energies above \ecross\ established the upward dispersion but were not adequate to resolve the symmetry.

In the present study, we return to the $x=0.04$ sample and probe the spin excitations with the time-of-flight chopper spectrometer SEQUOIA at the Spallation Neutron Source (SNS).  Beyond reproducing the anisotropy of the low-energy dispersion, we demonstrate that the excitations above \ecross\ form a spin-wave-like cone centered on \qaf.  Thus, we confirm in an LSCO crystal the energy-dependent symmetry change of the spin excitations first detected in YBCO,\cite{stoc04,stoc05,hink07,hink10} and we argue that the behavior is, therefore, compatible with stripe correlations.  We also present further triple-axis measurements that characterize the thermal evolution of the low-energy ($\lesssim10$~meV) magnetic spectral weight.

The rest of the paper is organized as follows.   The experimental methods are described in the next section.  The results are presented and analyzed in Sec.~\ref{sc:results}, while their significance is discussed in Sec.~\ref{sc:disc}.  The paper closes with a summary.

\section{Experimental Details}

The single crystal of La$_{1.96}$Sr$_{0.04}$CuO$_{4}$ used in this study is the same one used previously in Ref. \onlinecite{mats08}. The dimensions of the crystal, which was grown by the traveling solvent floating zone method, are $\sim6\,\phi\times25$~mm$^{3}$. As described in Ref. \onlinecite{mats08}, the crystal corresponds almost entirely to a single orthorhombic domain, which is vital to resolving the magnetic dispersion unambiguously.

The bulk of the inelastic scattering measurements were performed on the time-of-flight chopper spectrometer SEQUOIA \cite{sequoia06,sequoia10} installed at the SNS, Oak Ridge National Laboratory. Two incident energies, $E_i$, of 30 and 80 meV were used. The instrumental energy resolution is 1.4 ~meV (4.3 meV) at the elastic position for incident neutrons with an energy of 30 meV (80 meV).

The single crystal, with its $c$ axis aligned along the incident beam direction, was mounted in a closed-cycle $^4$He gas refrigerator.  For a given energy transfer $E=\hbar\omega$, the area detector maps out the intensity as a function of wave vector $(H,K,L_0)$, where $L_0$ is a constant that depends on $E$.  The magnetic correlations are known to be rather two-dimensional (2D), with no dependence on momentum transfer perpendicular to the CuO$_2$ layers. \cite{shir87}  As a consequence, this configuration allows the dispersion relations in the corresponding $(H,K)$ plane to be measured without rotating the sample. The $(H, K, L_0)$ and $(-H, K, L_0)$ data, which are equivalent, were summed in order to improve the statistics. The summed data were used to analyze the dispersion relations.

Additional low-energy excitation measurements were performed on triple-axis spectrometer TAS-2 installed at the JRR-3 facility of Japan Atomic Energy Agency. Neutrons with a fixed final energy, $E_f$, of 13.7 meV, together with a horizontal collimator sequence of guide-$80'$-$S$-$80'$-open, were used. The instrumental energy resolution is 1.5 meV at the elastic position. Pyrolytic graphite filters were used to suppress higher harmonics. The single crystal was oriented in the $(HK0)$ scattering plane and mounted in a closed-cycle $^4$He gas refrigerator.

\section{Results}
\label{sc:results}

\begin{figure}[t]
\includegraphics[width=8.5cm]{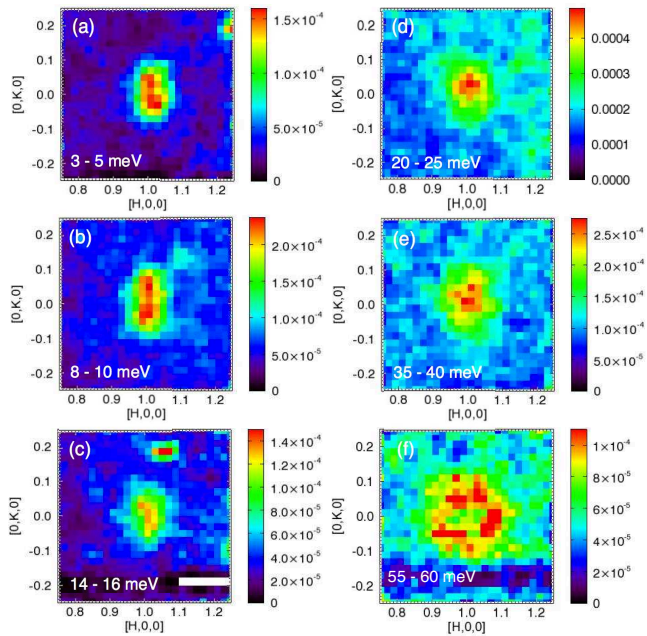}
\caption{(color online) Constant-energy slices around the antiferromagnetic wave vector $(1, 0, 0)$ at 10~K. The scattering intensity has been integrated over the energy range described in each panel. (a)-(c) and (d)-(f) are measured with $E_i=30$ and 80 meV, respectively.}
\label{2Dmap}
\end{figure}
Figure~\ref{2Dmap} shows contour plots of constant-energy slices of the inelastic neutron scattering spectra around ${\bf Q}_{\rm AF}= (1,0,0)$ between 3 meV and 60 meV in the $(HK)$ plane  measured at 10 K. The distribution of the magnetic signal, which at low energies is elongated anisotropicaly along the [010] direction of the modulation wave vector, becomes more isotropic about \qaf\ with increasing energy. At 55--60 meV, a ring-shaped excitation with an almost homogeneous intensity distribution, as one might expect from 2D isotropic spin-waves, is clearly observed.

\begin{figure}[t]
\includegraphics[width=8.5cm]{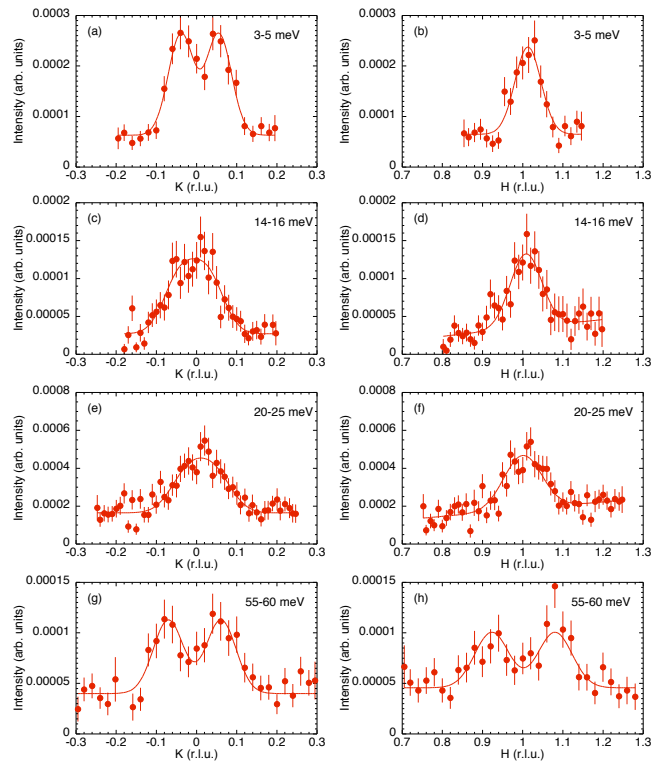}
\caption{(color online) Constant-energy cuts at 3--5, 14--16, 20--25, and 55--60 meV along $(0,K,0)$ and $(H,0,0)$ in La$_{1.96}$Sr$_{0.04}$CuO$_{4}$ at 10~K. The scattering intensity has been integrated over the energy range described in each pane. The scans along $(0,K,0)$ and $(H,0,0)$ correspond to those parallel and perpendicular to the incommensurate wave vector, respectively. (a)-(d) and (e)-(h) are measured with $E_i=30$ and 80 meV, respectively.  The lines through the data points are fitted gaussian peaks.}
\label{profiles}
\end{figure}
Figure~\ref{profiles} shows cuts along symmetry directions through the peaks of Fig.~\ref{2Dmap}, for energies between 3 meV and 60 meV. The panels on the left (right) correspond to a range of wave vectors running parallel (perpendicular) to the incommensurate modulation. At $\hbar\omega=3$--5 meV, the magnetic excitation peaks are observed at $(1, \pm\delta, 0)$, as previously reported.\cite{mats08} The peak splitting becomes smaller with increasing energy. Though the peak splitting is no longer resolved at $\hbar\omega=14$--16 meV, the peak width remains anisotropic, being larger along $K$.  In contrast, the peak shape is almost isotropic at $\hbar\omega=20$--25 meV. A two-peak structure is seen along both $H$ and $K$ at $\hbar\omega=55$--60 meV.  The lines through the data points are fitted gaussian peaks.

\begin{figure}
\includegraphics[width=8.5cm]{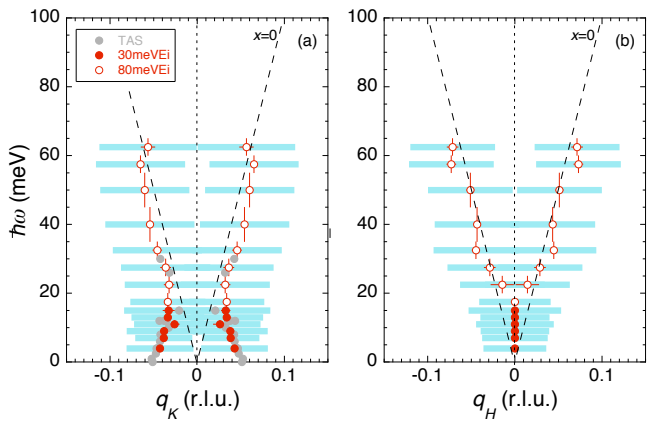}
\caption{(color online) Magnetic dispersion relation along $q_K$ (a) and $q_H$ (b) in La$_{1.96}$Sr$_{0.04}$CuO$_{4}$ at 10~K. The filled and open symbols represent the data points determined from the spectra measured with $E_i=30$ and 80 meV, respectively. The data measured previously using a triple-axis spectrometer\cite{mats08} are also plotted. The data plotted here were determined using the two Gaussians. Since in the dispersion along $q_H$ the scattering data below 20 meV was fitted using the one Gaussian, the data were plotted at $q_H=0$ (b). The fitted peak width (full-width-at-half-maximum of the peak) are shown with shaded horizontal bars. It is clearly seen that the overall peak width below 20 meV is much sharper along $q_H$ than $q_K$, although it is almost isotropic above 20 meV. The error bars along energy corresponds to the energy region where the scattering data is integrated.
The dashed lines indicate the spin wave dispersion in pure La$_2$CuO$_4$ with the spin wave velocity of 850 meV\,\AA.\cite{aepp89,cold01}
}
\label{dispersions}
\end{figure}

The fitted peak positions are indicated by circles in Fig.~\ref{dispersions}; the shaded horizontal bars represent the fitted peak widths. (Symmetry about \qaf\ was assumed in the fitting function.)  From the variation in the peak positions and widths, we estimate that $E_{\rm cross}=22\pm3$~meV; this is consistent with Ref.~\onlinecite{mats08} but with a reduced uncertainty.  For $E<E_{\rm cross}$, the peak widths are slightly broader along $q_K$ than along $q_H$, indicating some anisotropy even beyond the incommensurability.  Along $q_K$, the peaks disperse inward with increasing energy up to $\sim20$ meV and then change to an outward dispersion above $\sim30$ meV. On the other hand, along $q_H$, the excitations are mostly single peaked below \ecross, with the outward dispersion gradually becoming resolvable above $\sim$30 meV.  The magnetic dispersions above \ecross\  shown in Fig.~\ref{dispersions} are qualitatively consistent with those of commensurate spin-wave excitations in pure La$_2$CuO$_4$ \cite{aepp89}, although the slope, which corresponds to the spin-wave velocity, is slightly smaller in La$_{1.96}$Sr$_{0.04}$CuO$_{4}$.

Besides dispersion, it is also of interest to consider the energy dependence of the magnetic spectral weight.  We start by extracting the dynamical spin susceptibility $\chi''({\bf Q},\omega)$ from the measured scattering cross section using the relation\cite{hayd98b}
\begin{equation}
\frac{d^2\sigma}{d\Omega dE} = \frac{2(\gamma r_e)^2}{\pi g^2 \mu_B^2}\frac{k_f}{k_i}|f({\bf Q})|^2\frac{\chi''({\bf Q}, \hbar\omega)}{1-{\rm exp}(-\hbar \omega/kT)},
\nonumber
\label{intensity}
\end{equation}
where $(\gamma r_e)^2$=0.2905 barn sr$^{-1}\mu_B^{-2}$, $k_i$ and $k_f$ are the incident and final neutron wave vectors, and $f({\bf Q})$ is the magnetic form factor calculated for the Cu 3$d_{x^2-y^2}$ orbital. \cite{sham93}  The measured signal also depends on the size of the sample.  Conversion of the signal to absolute units was performed by properly normalizing to the elastic nuclear incoherent scattering from the sample.   A useful measure of the magnetic spectral weight is then given by the local susceptibility, defined by
\begin{equation}
  \chi''(\omega) = \int d{\bf Q} \,\chi''({\bf Q},\omega)\, \bigg/ \int d{\bf Q} .
\end{equation}
While one should, in principle, integrate over the entire Brillouin zone, in practice we integrate just the identifiable magnetic signal close to \qaf.

\begin{figure}
\includegraphics[width=8.0cm]{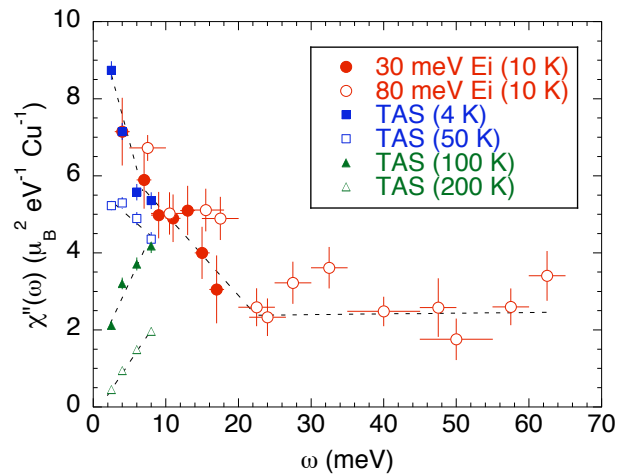}
\caption{(color online) Momentum-integrated $\chi''(\omega)$, representing the magnetic spectral weight, vs.~excitation energy. The horizontal error bar corresponds to the energy region where the data are integrated. The filled and open circles correspond to the data at 10 K analyzed using the data with $E_i=30$ and 80 meV, respectively.  The temperature dependence of the low energy $\chi''(\omega)$ ($\le8$ meV), measured on a triple-axis spectrometer, is also shown. The broken lines are guides to the eye.}
\label{I_E}
\end{figure}

The magnetic spectral weight as a function of energy is plotted in Fig.~\ref{I_E}, with the SEQUOIA data at 10~K represented by circles.  The distribution of spectral weight is consistent with results for LSCO with $x=0.05$,\cite{fuji12a} but it is intermediate between results for antiferromagnetic La$_2$CuO$_4$ (Ref.~\onlinecite{hayd96a}) and superconducting LSCO near optimum doping.\cite{chri04,vign07}   In the antiferromagnetic state, $\chi''(\omega)$ is constant in energy for energies less than the superexchange energy ($\approx143$~meV),\cite{head10} except for rather small energy gaps.\cite{kast98}  Of course, La$_2$CuO$_4$ also has significant weight in the antiferromagnetic Bragg peaks, which is at zero energy.  The holes doped into the planes in our LSCO $x=0.04$ sample frustrate the static order, and effectively push much of the associated spectral weight out to finite energy.  Thus, the pile up of weight at low frequency can be viewed as quasi-elastic scattering associated with the glassy order.  Above 20~meV, $\chi''(\omega)$ plateaus at a magnitude comparable to that in the antiferromagnet.\cite{hayd96a}

In the case of superconducting LSCO near optimal doping, a gap develops at low energy, with weight moving into a peak centered at 18~meV; a second peak in spectral weight occurs near $E_{\rm cross}\approx 45$~meV.\cite{vign07}   For our case of $x=0.04$, the quasi-static antiferromagnetic order would appear to be an obstacle to superconducting order.  

The temperature dependence of $\chi''(\omega)$ for $\hbar\omega\le8$~meV has been determined by triple-axis measurements; the results are indicated by filled and empty squares and triangles in Fig.~\ref{I_E}.  The results are in good agreement with the time-of-flight data at low temperature.   As temperature increases, the quasi-elastic peak decreases and disappears by 100~K, where the resistivity begins to develop a metallic temperature dependence.\cite{ando01}  The previous study\cite{mats08} showed that the incommensurability of the low-energy excitations also decreases with temperature.   The development of anisotropy in dc and low-frequency optical conductivities\cite{dumm03} is correlated with the growth in $\chi''(\omega)$ at low frequencies on cooling.

\section{Discussion}
\label{sc:disc}

We implied in the introduction that the study of LSCO with $x=0.04$ has relevance to stripe physics.
The incommensurate spin modulation is certainly compatible with spin stripes; however, a corresponding charge modulation has not been directly detected.  The possibility that the spin modulation corresponds to spiral order has been proposed.\cite{sush05,juri06,lusc07}  Given the disorder in the system, spiral correlations are likely present; however, a pure spiral has an instability to amplitude modulation in the presence of charge inhomogeneity.\cite{seib11}  Furthermore,
the drastic impact on antiferromagnetic ordering of a rather small density of doped holes suggests that the doped holes induce a strong frustration that is inconsistent with a uniform spiral order.  Indeed, the low-temperature optical conductivity of LSCO with $x=0.04$ is quite anisotropic, with a bigger gap along the $b$ axis, the direction of the spin modulation, and a reduced gap along the $a$ axis.\cite{dumm03}  Further evidence for glassy charge order comes from studies of resistance noise,\cite{raic08,raic11} and we note that the spin order occurs well below the temperature at which the in-plane resistivity begins an insulator-like upturn.\cite{ando01}  A recent model indicates how diagonally-oriented charge-stripe-segments, with associated vortex and antivortex spin textures in the antiferromagnetic background, can collectively produce a stripe-like texture that is compatible with experiment.\cite{seib13}  

Diagonal spin modulations have also been detected recently\cite{enok11,enok13} in the spin-glass regime of Bi$_{2+x}$Sr$_{2-x}$CuO$_{6+y}$; however, they are not unique to cuprates.  Diagonal stripe order has been observed widely in layered transition metal oxides doped with holes,\cite{ulbr12b} including nickelates,\cite{tran94a} cobaltates,\cite{cwik09} and manganites.\cite{ster96,ulbr11} The hourglass dispersion of magnetic excitations has been observed in both the cobaltate\cite{boot11} and manganite\cite{ulbr12a} systems.  For the cobaltate and manganite cases, the measured spectra are described fairly well by spin-only models when stripe disorder is taken into account.\cite{andr12,ulbr12a}  The magnetic excitations in \lsno\ do not exhibit a full hourglass spectrum\cite{bour03,boot03,woo05,free05}; nevertheless, the observed dispersions are reproduced fairly will by spin-wave theory.

While the hourglass spectrum is a common feature, there are several differences between cuprates and the other transition-metal oxides.  First of all, we expect that inelastic measurements on single-domain samples of the other stripe systems would find two-fold rotational symmetry at all energies.  The four-fold symmetry for $E>E_{\rm cross}$, first detected in YBCO\cite{stoc04,stoc05,hink07,hink10} and now confirmed in LSCO, is unusual and challenging to understand.  Secondly, incommensurate spin modulations appear in cuprates at very low doping, whereas they only become apparent in the other systems at substantially higher hole concentrations.  For example, in nickelates stripe order has only been detected\cite{sach95,yosh00} for hole concentrations $\gtrsim0.14$; this behavior is correlated with the degree of carrier localization.  The doped holes in cuprates exhibit less localization, and hence can develop spatial correlations at a rather low density.

The final difference among materials is superconductivity, which occurs only in the cuprates.  To be accurate, our LSCO $x=0.04$ does not exhibit superconducting order; superconductivity only appears for $x\gtrsim0.055$.  It is intriguing to note, however, that a scaling analysis of resistance as a function of doping in an electrolytically-tuned LSCO thin film indicates that the superconductor-to-insulator transition involves the localization of pairs.\cite{boll11}  Furthermore, magnetoresistance\cite{shi13} and magnetization\cite{li07b} studies suggest the presence of superconducting fluctuating on the insulating side of the transition.  On the theory front, Scalapino and White\cite{scal12b} have argued that charge stripes form from paired holes.   While they have in mind bond-parallel stripes at higher doping, it is interesting that Seibold {\it et al.}\cite{seib13} have integral numbers of hole pairs in the diagonal ``ferronematic'' segments that they propose for modeling the spin-glass phase of LSCO.

Could the existence of pairing play a role in the restoration of rotational symmetry at $E>E_{\rm cross}$?  In the absence of any theory, we put forward a wild speculation.  A hole-rich stripe segment will contain spin degrees of freedom.   If the holes in the segment are paired, then one might expect the spins to form singlet pairs.  The $\pi$ phase shift in the antiferromagnetic background on crossing a charge-stripe segment can minimize the correlations between the spin background and the stripe segment, thus providing some protection to the singlets associated with the hole pairs; it is this correlation, pinned to the anisotropic lattice potential, that underlies the uniaxial spin-stripe order.   The protection of pairs is important for energies below the singlet-triplet gap, which would be comparable to \ecross.  Above \ecross, the antiferromagnetic spin excitations no longer have a twist in them.  A spin flip at any position will effectively create a local triplet that can propagate in an antiferromagnetic background, so that one might expect to recover four-fold symmetry.   We repeat that this is a purely speculative scenario which we offer in the hope of motivating further research.

\section{Summary}
We have performed quantitative analysis of the magnetic dispersions up to 65 meV in La$_{1.96}$Sr$_{0.04}$CuO$_{4}$. It is found that the anisotropic excitations at low energies, in which outward dispersing branches are missing, gradually change to isotropic ones with a conical dispersion relative to \qaf, as in the parent compound La$_2$CuO$_{4}$. The low-energy $\chi''(\omega)$ develops gradually with decreasing temperature. Interestingly, these observations are similar to behavior first detected in underdoped \ybco.\cite{stoc04,stoc05,hink07,hink10}  The surprising restoration of symmetry at high energy is in need of a theoretical explanation.

\begin{acknowledgments}
This research at ORNL's Spallation Neutron Source was sponsored by the Scientific User Facilities Division, Office of Basic Energy Sciences, U.S. Department of Energy. The work at BNL was supported by the U.S. DOE's Office of Basic Energy Sciences, Division of Materials Sciences and Engineering, under Contract No. DE-AC02-98CH10886.
\end{acknowledgments}


%

\end{document}